# Recent Developments in Immune Network Theory including a concept for an HIV Vaccine

Geoffrey W. Hoffmann

Network Immunology Inc.
3311 Quesnel Drive
Vancouver, B.C.
Canada V6S 1Z7

and

Department of Physics and Astronomy
University of British Columbia
Vancouver, B.C.
Canada V6T 1Z1

Abstract

The symmetrical network theory is a framework for understanding the immune system, that dates back to the mid 1970s. The symmetrical network theory is based on symmetrical stimulatory, inhibitory and killing interactions between clones that are specific for each other. Previous papers described roles for helper and suppressor T cells in regulating immune responses[1,2] and a model for HIV pathogenesis.[3] This paper extends the theory to account for regulatory T cells that include three types of suppressor cells[4] called Ts1, Ts2 and Ts3, and two types of helper cells[5] called Th1 and Th2. The theory leads to a concept for an HIV vaccine, namely a reagent commonly known as IVIG, to be administered in small amounts in an immunogenic form via an immunogenic route. Predictions are made for experiments in mice and macaque monkeys.

E-mail: hoffmann@networkimmunologyinc.com

**Introduction**

The symmetrical network theory is a framework for understanding the adaptive immune system.[1-3] The symmetrical network theory involves three types of interactions, namely stimulatory, inhibitory and killing interactions. All of these are symmetrical. If the variable region of an antibody recognizes a complementary variable region, the latter also recognizes the former. This aspect of the theory is called "first symmetry". Stimulatory interactions involve cross-linking of the specific receptors of lymphocytes. The lymphocytes that make antibodies are called B cells. Symmetry in stimulation means for example that if antibodies made by a B cell "Y" are able to cross-link the receptors on a B cell "Z", thus stimulating the cell, then the antibodies made by the B cell "Z" are likewise able to cross-link the receptors of the B cells "Y". Antibodies have a molecular weight of about 150,000 daltons or more, and are able to cross-link receptors because each antibody has two or more V regions. The cross-linking of surface immunoglobulin is a necessary but not sufficient condition for switching on B cells to secrete antibody. Cross-linking alone stimulates B cells to proliferate. Anti-Ig is stimulatory, $F(ab)_2$ fragments of anti-Ig are likewise stimulatory, while Fab fragments of anti-Ig do not stimulate, and Fab fragments of anti-Ig together with anti-Fab antibodies are stimulatory.[4]

The lymphocytes that regulate the production of antibodies by B cells are called T cells. T cells secrete molecules that include antigen-specific T cell factors. In the symmetrical network theory these molecules play a key regulatory role, and we need a shorter word for them, so I will call them tabs. Tabs were discovered by Nelson in 1970[7], and were studied by many immunologists in the 1970s and 1980s. Tabs have a molecular weight of about 50,000 daltons, from which it is plausible that, in contrast to antibodies, they have only one V region. Then tabs are able to block complementary V regions, but not cross-link them. Hence they can have an inhibitory function. In the regulation of the immune response to an antigen X, there are tabs with V regions that bind to X ("anti-X tabs") and tabs that bind to anti-X V regions ("anti-anti-X tabs"). The theory predicted the existence of and a regulatory role for anti-anti-X tabs, and this prediction has been confirmed.[8]

Killing by antibodies is likewise symmetrical, as demonstrated by Anwyl Cooper-Willis and myself.[9] The experiment involved Fab fragments of two mutually specific IgM monoclonal antibodies, HPC-M2 and B36-82, that were covalently coupled to two sets of red blood cells (RBC). The RBC-HPC-M2-Fab cells were lysed by B36-82 antibodies, and the RBC-B36-82-Fab cells were lysed by HPC-M2 antibodies. This experiment confirmed a simple, basic postulate of the symmetrical network theory. In fact the theory predicted the result.

A recurring theme in the symmetrical network theory is co-selection. Co-selection means mutual positive selection of members of two diverse populations, such that selection of members within one population is dependent on interactions with one or more members within the other population. Helper T cells are selected to have some affinity to MHC molecules called MHC class II. In the symmetrical network theory suppressor T cells are selected that have V regions with affinity for helper T cell V regions. There is mutual stimulation between the helper T cells and the suppressor T cells. Hence helper T cells are selected that not only have affinity for MHC class II, but also for suppressor T cell V regions. The suppressor T cells that are selected have complementarity to as many helper T cell V regions as possible. This constraint results in the emergence of a mutually stabilizing set of helper T cells and suppressor T cells. The number of ways for suppressor T cells to have V regions that interact with a large number of different helper T cells is limited, and the selective process results in the suppressor T cells expressing serologically detectable determinants, that are indirectly influenced by MHC class II.[10] Since the helper T cells are anti-MHC class II and the suppressors are anti-anti-MHC class II, we say that the suppressor V regions are an "internal image" of MHC class II. They are an image of MHC class II in the context of the repertoire of helper T cell V regions. I have used a Mongolian yurt as an analogy for the mutually stabilizing interactions between the helper cells and the suppressor cells. The suppressor cells are the centre-pole, and the helper cells are the canvas. The centre-pole holds up the canvas and the canvas holds up the centre-pole. The emergent centre-pole is the central regulating element of the system. This model explains the highly controversial phenomenon of I-J, in which suppressor T cells express serological determinants that map to within the MHC, without any I-J gene or genes being present in the MHC.[3]

The theory includes a model for HIV pathogenesis, in which it was postulated that HIV preferentially infects HIV-specific T cells.[3] As a result, there is co-selection also of helper T cells and HIV. The criterion for selection of HIV variants is then the same as the criterion for the selection of suppressor T cells, namely to have shapes (specific receptors) with complementarity to as many helper T cell V regions as possible. Therefore with time the HIV variants selected *in vivo* look more and more like the V regions of the suppressor T cell population. The immune response against HIV then becomes an immune response also against the suppressor T cells. This attack on the central regulating element of the system leads to autoimmunity and the collapse of the system. Eight years after this model was published, the postulate that HIV preferentially infects HIV-specific T cells was shown to be correct.[11]

In this paper I extend the scope of the symmetrical network theory by incorporating three types of suppressor T cells, namely Ts1, Ts2, and Ts3, and two kinds of helper T cells called Th1 and Th2. The model leads to a postulated role for serum IgG in the maintenance of self tolerance. The model is then further extended to include both IgM secreting B cells and IgG secreting B cells. The model provides a rationale for the fact that HIV is not as highly infectious as many other viruses.

I will first review some of the facts about Ts1, Ts2, Ts3, Th1, Th2 and serum IgG. I will then present an extended version of the symmetrical network theory model that explains the phenomena in that context. I will finally describe how the theory leads to a concept for a vaccine for the prevention of infection with HIV.

**The suppressor T cells Ts1, Ts2 and Ts3**

For a given antigen X, some strains of mice make anti-X antibodies that have consistent V regions, in that they bind to certain anti-anti-X antibodies called antiidiotypic antibodies. V regions that bind to antiidiotypic antibodies express the idiotype defined by the antiidiotypic reagent.

Evidence was obtained in the early 1980s for the existence of three kinds of suppressor T cell, that were known as Ts1, Ts2 and Ts3 cells. Some investigators gave them the names "suppressor inducer" (Ts1), "suppressor amplifier" (or "acceptor", Ts2), and "suppressor effector" (Ts3).[4] The Ts1 cells are likely the same cells that have more recently been called Treg cells.[12,13] In an immune response antigen-specific Ts1 cells induce Ts2 cells, and Ts2 induce Ts3 cells. Ts3 is typically a population that has V regions that are similar to the V regions present on Ts1 cells, as defined by binding to antiidiotypic antibodies, but are different in that their V regions do not bind to the antigen.[4] If T cells induced by Ts2 do bind antigen, they are part of the Ts1 population by definition, and in some systems only Ts1 and Ts2 but not Ts3 have been described. Antigen-related Ts2 cells are anti-anti-X, interact with both Ts1 and Ts3 cells, and are accordingly antiidiotypic. The presence of idiotypic and antiidiotypic suppressor T cells in the Ts1, Ts2, Ts3 cascade is unambiguous evidence of idiotypic network regulation of the immune system.

Tabs are involved. The antigen acts on Ts1 cells, which secrete tabs called TsF1, that act on (induce) Ts2, and these secrete tabs called TsF2, that then in turn act on (induce) Ts3, that secrete TsF3. As is the case for the corresponding T cells, TsF1 and TsF3 typically express the idiotype defined by binding to antiidiotypic antibodies, while the TsF2 tabs express the corresponding antiidiotype.

**Th1 and Th2 helper T cells**

Tomio Tada and his collaborators found evidence for two types of helper T cells, one of which they called Th1, and one that they called Th2. The Th2 cells express an antigenic determinant that is expressed on suppressor T cells, and which is not present on Th1 cells.[5] These two kinds of helper T cells were defined in *in vivo* experiments, and are not to be confused with Th1 and Th2 cell lines that were later defined on the basis of the cytokines the T cells produce. Th1 cells provide "cognate" help, meaning that the B cells are specific for a small molecule called a hapten, and the helper T cells are specific for larger molecules called carriers, with the haptens being covalently coupled to carrier molecules. In this kind of immune response a T cell, a B cell and the antigen need to be located close to each other. Th2 cells provide

"non-cognate" help, meaning that an immune response can occur with the hapten coupled to a second carrier, that is different from the one to which the hapten is coupled. Non-cognate help reflects action at a distance, that is mediated by non-specific mediators called lymphokines. The Th2 cells in the experiment of Tada et al. express a serologically detectable determinant that is present on Ts2 cells, while the Th1 cells do not express this determinant.

**IgG dimers in pooled sera**

When many human sera are pooled and then examined under an electron microscope, dimers of IgG molecules are observed.[14] The dimers can be seen in the electron microscope images to be bound to each other via V-V interactions. Very few such dimers are seen in IgG from a single serum. This finding fits with the basic ideas of the symmetrical network theory in the context of the assumption that the dimers are pairs of IgG molecules with a high affinity for each other. The IgG specificities in a single serum include those for which the immune system is in the immune state. IgG molecules kill cells with high affinity complementary specificities, such that clones with complementary high affinity V regions are effectively eliminated within a single person. When serum antibodies from one individual are combined with IgG from other sera, the IgG idiotypes in those other sera include different ones, and hence there are complementary clones that have not been eliminated. The more sera that are combined, the higher the fraction of IgGs that form dimers. While a plasma sample from a single individual contains less than one percent of the IgG molecules as IgG-IgG (idiotype-antiidiotype) complexes, pools of IgG from a very large number of donors can contain as much as 40% of the IgG being present as dimers.

**A model incorporating Ts1, Ts2, Ts3, Th1, Th2 and serum IgG**

While Ts1, Ts2, Ts3, Th1 and Th2 cells have been defined experimentally in the context of particular foreign antigens, it is also possible to define broader classes of Ts1, Ts2, Ts3, Th1 and Th2 cells, related to MHC class II as a key self antigen. Each of the antigen-related Ts1, Ts2, Ts3, Th1 and Th2 cells are then plausibly subsets of MHC class II defined sets.

A co-selection model that incorporates MHC class II related Ts1, Ts2, Ts3, Th1, Th2 and serum IgG is shown in Figure 1. This is a generalization of the simpler co-selection model of reference 3, which included only Th1 and Ts2 as shown here. As stated above, co-selection of two populations means a combination of members of one population being selected by interactions with members of a second population, and members in the second population being selected by interactions with members in the first population. The Th1 population is selected by interactions with MHC class II and with Ts2 cells. There is co-selection of Th1 cells and Ts2 cells, with the Ts2 cells being the centre-pole as previously described.[3] However, in the more detailed model of Figure 1 there is also co-selection of Ts1 and Ts2 cells and co-selection of Ts2 and Ts3 cells. The Ts1 population is shown being selected by interactions with MHC class II bearing cells, and being co-selected with Th2 cells and serum IgG. Ts3 cells are co-selected with Ts2 cells, Th2 cells and serum IgG. "IgG" in this diagram stands for both serum IgG and IgG secreting B cells. The concept of co-selection applies strictly to the IgG producing B cells, since IgG molecules cannot be stimulated to proliferate. Nevertheless serum IgG may be more important in stimulating Ts1 and Ts3 cells than the B cells, since IgG is typically present at the high concentration of about 10mg/ml in serum, and being divalent means it is able to cross-link complementary receptors. B cells diffuse much more slowly than IgG, and actual B cell mediated stimulation of cells with complementary receptors would require cell-cell contact. Hence IgG mediated stimulation is plausibly more important.

Figure 1 is a steady state model that shows how the various populations are related to MHC class II. Within each of the cellular populations shown there are both antigen-specific and antiidiotypic cells for any antigen, and the serum IgG likewise includes both antigen-specific and antiidiotypic antibodies for any antigen.

In this model both Th2 and Ts2 cells are anti-anti-MHC class II, but Th2 cells differ from Ts2 in that they are co-selected by Ts1 and Ts3, while Ts2 cells are co-selected by Th1, Ts1 and Ts3. Since Ts2 cells are co-selected with three other populations, they are the most tightly regulated cells in the system. This is consistent with them being the centre-pole of the network.

In the model the interactions of Th2 cells with both Ts1 and Ts3 means that Th2 cells are more tightly regulated than are Th1 cells. On the other hand, when Th2 cells are stimulated by an antigen, this constitutes a more profound perturbation close to the heart of the system, resulting in the release of lymphokines that can provide help over a greater distance. This is consistent with the activation of Th2 cells providing "non-cognate help", as reported in the experiment of Tada et al.[5]

The model of Figure 1 involves the serum IgG repertoire being sharply constrained directly by interactions with Ts1 and Ts3, and by indirect interactions with the other populations shown. Ts1 and Ts3 cells are in turn both reciprocally regulated by the Ts2 centre-pole. Since serum IgG is similar to Ts2 in that it is anti-anti-MHC class II, it expresses idiotypic determinants similar to those of Ts2. This commonality means the V regions of serum IgG are far from random; they are all both anti-Ts1 and anti-Ts3, and both anti-anti-MHC class II. This constraint means that serum IgG molecules are less diverse than normally imagined; it means that serum IgG is a quasi-species. This means serum IgG is a diverse set of antibodies, that nevertheless all have something in common in their V regions due to the selection constraints. This property explains the surprising fact that IgG from a large number of donors has to be combined in order to obtain many high affinity IgG dimers. The idea that serum IgG is a quasi-species is a stark departure from traditional thinking, since the clonal selection theory involves the ability of an individual to make antibodies with a practically unlimited range of specificities.

The model of Figure 1 refers to the repertoire of serum IgG antibodies and of the B cells that are co-selected to secrete IgG. The concept that serum IgG antibodies have V regions with idiotypic determinants that resemble the idiotypic determinants on Ts2 cells leads to the surprising prediction that immunization of a mouse with IgG from another mouse, that differs only in its MHC genes (an MHC congenic strain), will result in the production of strain-specific antibodies that bind to Ts2 V regions.

**Igh restriction in suppressor cell interactions**

In some systems the interactions of Ts cells and factors are restricted by antibody V region genes, namely the heavy chain locus called Igh. This phenomenon is called Igh (immunoglobulin heavy chain) restriction. In the p-azobenzenearsonate system of Benacerraf and colleagues, the action of Ts1, Ts2 and Ts3 are all restricted by immunoglobulin heavy chain genes.[15] For example, antigen-specific suppressor tabs from BALB/c mice, which have the H-2$^d$, Igh-1$^a$ genotype, can suppress T cell responses in BALB/c mice but not in C.AL-20 mice, that are H-2$^d$, Igh-1$^d$. Similarly, antigen-specific suppressor tabs from C.AL-20 mice can suppress T cell responses in C.AL-20 but not in BALB/c mice. The repertoire of heavy chain genes (Igh genes) are both self antigens that affect the selection of regulatory cells, and also have a direct structural impact on the IgG repertoire. In our model serum IgG exerts a stimulatory role on Ts1 and Ts3, and indirectly also on Ts2, and thus plays an important role in the selection of those repertoires. Then the Ts1 repertoire and the Ts3 repertoire influence also the Ts2 repertoire, and serum IgG indirectly influences also the Ts2 centre-pole repertoire. By first symmetry, these influences are reciprocal, so that the serum IgG repertoire is also profoundly influenced by the Ts1, Ts2 and Ts3 repertoires. As a result of all this, T cells and tabs from mice with different Igh genes have not been selected *in vivo* to interact with each other, and suppression does not occur with mismatched populations.

**Immune responses: interplay between T cell and B cell repertoires**

The next step is to describe how the model of Figure 1 is consistent with the stable steady states for specific antigens of the symmetrical network theory,[1] together with the process of switching from a virgin state to an immune state or from a virgin state to a suppressed state for an antigen.[2]

Figure 1 shows only one "principal" axis in shape space, that can be considered to be defined by the anti-MHC class II specificity of Th1 cells of an animal or person, and the anti-anti-MHC class II specificity of the corresponding Ts2 cells. I will call this the host $\alpha$MHC class II/$\alpha\alpha$MHC class II shape space axis, where $\alpha$ is an abbreviation for anti. When an immune response to an antigen X takes place there is a change in the system with respect to another shape space axis, namely an axis defined by the αX receptors of antigen-specific lymphocytes and the ααX receptors of corresponding antiidiotypic lymphocytes. As already mentioned, for the αX/ααX shape space axis there are both αX and ααX lymphocytes and antibodies among each of the populations shown in Figure 1. The response to the antigen X is governed by the fact that the Th1 cells are the least tightly regulated and hence the most sensitive to stimulation by antigen, and antigen-specific cells are stimulated before the corresponding antiidiotypic cells. If the stimulus, say antigen X, results in the activation of non-specific accessory cells, and αX B cells differentiate to produce αX IgG, there is a change in the system with respect to the αX/ααX shape space axis associated with the antigen, and there is memory (immunity) associated with that change. The system then reverts to close to its original position with respect to the host αMHC class II/ααMHC class II shape space axis, while having a new position with respect to the αX/ααX shape space axis, including changes in the levels of antigen-specific B cells and antigen-specific IgG.

Early in the development of the symmetrical network theory the activation of non-specific accessory cells was postulated to play a role in immune responses.[2] If non-specific accessory cells are not activated, there can nevertheless be changes in the Ts populations, and antigen-specific tolerance may be induced, without any production of antigen-specific IgG antibodies. Memory can be associated not only with immunity, but also with antigen-specific tolerance. The suppressed state of the symmetrical network theory is characterized by elevated and mutually stabilizing populations of antigen-specific and antiidiotypic T cells.[2] Such changes in the T cell repertoire necessarily result in a change in the Ts2 centre-pole of the system. The change in the Ts2 population results in changes in the other populations, including the IgG repertoire. An increase in antigen-specific and antiidiotypic T cells (induction of the suppressed state) can be expected to correlate with a decrease in the level of antigen-specific serum IgG. The level of this IgG is predicted to be lower than the level for an immune system that is in the virgin state for the antigen, since the elevated Ts and TsF levels result in the inhibition of B cells that secret the antigen-specific IgG.

While the T cell V region repertoire of an individual influences the B cell V region repertoire, and the influence is reciprocal, the immune system is regulated primarily by T cells, that are more sensitive to antigen than are B cells. This is consistent with the model of Figure 1, in which the Th1 cells are ascribed lower network connectivity than the Ts2 cells, and are the least inhibited for proliferation in response to antigen.

The presence in the model of serum IgG at the heart of the self-stabilized system is consistent with the concept that serum IgG is a quasi-species, that mimics the Ts2 central regulating element. It is also consistent with the concept that there is immunological memory for immune responses that exhibit changes in the serum IgG population, with associated changes in the Ts2 repertoire.

The two major classes of antibody in serum are IgM and IgG. Interaction matrices between IgM monoclonal antibodies derived from neonates have been measured[16] and found to have a connectance (fraction of non-zero terms) of about 0.2. IgM responses are not regulated by T cells to the same extent that IgG responses are, and there is no memory associated with immune responses that are solely IgM. In the symmetrical network theory IgM plays a role in the virgin state, which is a stable steady state involving many specificities and symmetrical killing between mutually specific clones. Mathematical modelling of the postulated IgM interactions has shown that for a given set of IgM secreting clones there is a unique stable steady state.[17] The results obtained with that mathematical model are consistent with the finding that no memory is associated with immune responses that include only IgM without an IgG component.

**A balance in self antigens between shapes and complementary shapes?**

The immune system is a highly sensitive system that can be modulated by very small amounts of antigens and antibodies. Experiments in rats show that the specific response of the system to a particular antigen can be significantly decreased by prior injections of antigen as low as picograms or even less.[18] A response to antigen consisting of antibodies with a particular idiotype can be suppressed by an injection of mice with 10 to 100ng of antiidiotypic antibody.[19] The injection of mice with nanogram amounts of monoclonal IgM antibody can induce the production of antibodies of the same specificity.[20] It would seem that such manipulations of the immune system can make a marked difference to the state of the system only if it is normally precisely balanced, such that very small perturbations can shift the state of the system significantly. This may be because a balance between each shape and other shapes, that are complementary to the shape, is a basic feature of the repertoire of self antigens that impact on the V regions of the immune system. We can call this the "balanced proteome hypothesis." This hypothetical balance in the proteome of self antigens may be enhanced by the steady state response of the immune system to the repertoire of self antigens, with any imbalance in the self antigen repertoire being dynamically compensated by adjustments in V region repertoires.

There are many soluble self antigens that have complementary cell surface receptors to which they bind. These are examples of shapes and corresponding complementary shapes. The stimulation of the T cell V region repertoire by a soluble self antigen may be more or less balanced by stimulation exerted by the corresponding complementary cell surface receptor.

There is some cross-reactivity between mouse MHC class II molecules and human MHC class II molecules,[21] which can also be understood in this context. The complete mouse and human genomes have recently been found to be surprisingly similar. In the context of evolutionary time, MHC molecules are selected in the context of the complete set of all the self antigens, and according to the balanced proteome hypothesis, the environment for the evolutionary selection of mouse MHC molecules is similar to the environment for the evolutionary selection of human MHC molecules.

In the context of the balanced proteome hypothesis, the addition of a new gene to the genome would then create a viable organism only if there is not a significant disruption of a hypothetical balance between shapes and complementary shapes. The maintenance of that balance may include the spectrum of both germ line and selected somatically derived V regions, which are or become part of the set of self antigens. This idea may be relevant for understanding a V region transgenic mouse experiment by Weaver et al. The mouse expressed IgG with the same idiotype as the idiotype of the transgene, but using different genes.[22]

**On a balanced set of self antigens and linkage disequilibrium**

The major histocompatability complex in humans is called HLA (human leukocyte antigens). A much studied phenomenon is linkage disequilibrium for HLA genes. This means that there is a tendency for particular HLA alleles to remain associated with each other. In the context of the idea that self antigens are a balanced set of shapes and corresponding complementary shapes, we can interpret HLA linkage disequilibrium in terms of some HLA genes combining with others to give more balanced stimulation of the T cell idiotypic network than other combinations would do. The precise shape of any given MHC molecule would not matter very much according to this view, rather the important thing is that the combination of all the MHC antigens, including MHC class I and MHC class II, results in an optimally balanced proteome. This constraint is not very restricting, and there would be many ways in which it can be satisfied. Hence we would have a simple explanation for the high polymorphism of MHC genes, that is an alternative to the conventional explanation, namely that MHC diversity protects against a species being decimated by a particular pathogen.

**A model with all the self antigens, and IgG and IgM producing B cells**

The model of Figure 1 can be extended to explicitly include all the self antigens, IgM producing B cells, denoted by B1, and IgG producing B cells, denoted by B2, as shown in Figure 2. The figure includes three level of lymphocytes, with level 1 being selected to have some anti-self affinity, level 2 being anti-anti-self, and level 3 being anti-anti-anti-self. IgM secreting B1 cells are shown as being selected solely by Ts2 cells, and as a result they are not as tightly regulated as IgG secreting B2 cells. The expressed serum concentrations of IgM and IgG of all specificities depend on the B1 and B2 repertoires and also on the various T cell repertoires.

In the model of Figure 2 MHC class II is replaced by all the self antigens, since we know that small changes in the repertoire of self antigens can result in significant changes in idiotypic repertoires. The self antigens of particular consequence may include proteins that are not polymorphic such as fetuin (a protein present at a high concentration in serum early in development), albumin, and complement components that just happen to be encoded within the MHC. All of these and also self antigens that are present in lower amounts would then play a role in the selection of the Th1 population, and hence the selection of the emergent Ts2 centre-pole. The effect of proteins that are not polymorphic on the Th1 and Ts2 repertoires would not be evident in the types of experiments that demonstrate the effect of polymorphic MHC class II on these repertoires.

While Figure 2 shows all the self antigens impacting on the Th1 population, MHC class II plays a particularly important role in this respect. The Ts2 population remains the central regulating element of the immune system, and it is constrained to be a homogeneous population primarily via co-selection with Th1 cells, the repertoire of which is influenced also by the repertoire of all the self antigens. Figure 2 is furthermore a model that shows in some detail how the repertoire of self antigens may interact with Th1 and Ts1 populations, how these may interact with the Ts2, Th2 and B2 populations, and how they in turn may interact with B1 and Ts3 populations. The figure illustrates the set of interactions with respect to an anti-self/anti-anti-self shape space axis.

**T-independent immune responses**

Antigens that induce an immune response that is solely IgM are typically "T-independent" antigens, and are typically multivalent, flexible molecules that efficiently cross-link specific receptors on lymphocytes. A T-independent antigen stimulates a broad spectrum of diverse idiotypes with a low average affinity at the level of the binding of a single epitope to a receptor. In the model the stimulated clones are so diverse that the antigen is eliminated by IgM before there is much co-selection of antigen-specific and antiidiotypic T cells.

**T-dependent immune responses and the Oudin-Cazenave paradox**

T-dependent antigens on the other hand include proteins that do not cross-link specific lymphocyte receptors as efficiently as is the case for T-independent antigens. For a T-dependent antigen X, the immune response involves co-selection of antigen-specific and antiidiotypic T cells. These are anti-X and anti-anti-X, or "$\alpha X$" and "$\alpha\alpha X$" respectively. This co-selection process is exactly analogous to the co-selection process for the self antigen MHC class II and the Th1 and Ts2 cells. Co-selection of the $\alpha X$ and $\alpha\alpha X$ T cell populations is a highly non-linear selection process, that can lead to the emergence of a characteristic, sharply defined antiidiotype in the ααX T cell population. This emergent population is likely just as important for the stimulation of αX B cells as the antigen itself. The ααX T cell V regions would also stimulate αααX B cells, that express the same idiotype as the αX B cells (as defined by an antiidiotypic reagent), but with V regions that do not bind to X. Hence this co-selection model explains the Oudin-Cazenave paradox, in which immune responses to an antigen X include antibodies that do not bind to X, and yet have the same idiotype as the αX antibodies.[23]

**Serum IgG may mediate self tolerance**

MHC class II molecules certainly have a strong impact on the repertoire of helper T cells. Less is known about the impact that less polymorphic molecules may have, including for example the serum proteins fetuin (a protein present at a high concentration in serum early in development), albumin and complement components. The fact that some complement genes are encoded within the MHC suggests that complement proteins could also influence the T cell repertoire significantly. It is plausible that Th1 and Ts1 cells are more generally anti-self, rather than solely anti-MHC class II, and that the Ts2 cells, Th2 cells and serum IgG are then accordingly more generally anti-anti-self, rather than solely anti-anti-MHC class II.

In the model of Figure 2 there is a symmetry in the T cell repertoire in the sense that it includes both anti-self and anti-anti-self regulatory T cells. Th1 and Ts1 cells are biased to be anti-self, while Th2 cells and Ts2 cells are anti-anti-self. On the other hand the selection of serum IgG molecules through their interactions with Ts1 and Ts3 cells results in the serum IgG having only anti-anti-self specificity and no anti-self specificity. This one-sidedness accounts for the low fraction of dimers that are formed when IgG from only a small number of donors is pooled. The anti-anti-self serum IgG molecules kill cells that have receptors with a high affinity for self, that is, anti-self lymphocytes. Thus this model describes how the emergent, selected repertoire of serum IgG deletes self-specific clones, and may therefore be an important mediator of self tolerance.

**Offense and defense**

As for countries fighting wars, the immune system is capable of both offense and defense in this model. Each IgG immune response includes an anti-foreign component (offense) and an anti-anti-self component (defense). The anti-foreign component eliminates the antigen, while the anti-anti-self component inhibits any of the anti-foreign cells that are also anti-self. The anti-anti-self response results from the selection of a new set of Th1 cells, with the emergence of a new set of co-selected Ts2 cells. So after each immune response that has long term memory, there is a slightly different centre-pole from the one that preceded the immune response. This change inevitably results in changes in the Ts1 and Ts3 populations, with the further result that there are changes also in the anti-anti-self IgG producing B2 cells.

**IgM and HIV**

In our theory of HIV pathogenesis, HIV evolves in infected people to mimic the receptors of the Ts2 central regulating element of the immune system.[3] People who become infected with HIV go through a phase of synthesizing large amounts of HIV, yet the amount of transmission of the virus to healthy uninfected individuals is low, compared for example with the infectivity of measles for people who have not been immunized against measles. Figure 2 provides a rationale for this. In the model the IgM repertoire is selected to have complementarity to Ts2, and since HIV is systematically selected *in vivo* to resemble Ts2, the IgM repertoire as a whole has complementarity also to HIV. Hence HIV is normally quickly cleared by IgM, without seroconversion to the production of anti-HIV IgG antibodies.

**MHC restriction of V-V interactions in serum IgG**

We recently discovered that purified IgG from mice of a given H-2 haplotype binds rapidly in an ELISA assay to purified IgG from mice with the same MHC haplotype, but not to purified IgG of a different MHC haplotype.[24] This is most simply interpreted in the context of our model as indicating that normal IgG has both anti-anti-self and anti-anti-anti-self specificity. The anti-anti-anti-self specificity is an added twist, that was not predicted by the model. It is however consistent with the MHC having a profound influence on the serum IgG repertoire, and can be explained as being the result of IgG clones being selected by a combination of anti-self and anti-anti-self tabs, that are present on the surface of non-specific accessory cells.[7] The affinity of this MHC restricted V-V binding as seen in the ELISA assay may be much lower than that for the dimers seen in pools of IgG from many unrelated donors.[14]

**A vaccine for the prevention of HIV infection**

The main difficulty for developing vaccines for the prevention of infection with HIV has been the enormous diversity of the virus. I here describe a candidate vaccine that addresses this problem partly by employing an equally diverse reagent, that has an important similarity to HIV. HIV evolves to resemble Ts2 V regions in the human population, and due to the co-selection processes described in this paper, the spectrum of Ts2 V regions in humans have an important similarity to the spectrum of IgG V regions in humans. I propose that a vaccine containing a reagent commonly known as IVIG can be used for the prevention of infection with HIV. IVIG is an acronym for intravenous immunoglobulin. IVIG is obtained from the pooled plasma of a large number of HIV negative blood donors. The postulated HIV vaccine consists of IVIG given in immunogenic form, namely with an adjuvant such as alum, via an immunogenic route, for example intra-muscularly. IVIG is available commercially, and very small amounts of it are expected to suffice for vaccination against HIV.

The vaccine has two modes of action, which have to do with the fact that the IgG of IVIG includes both shapes that are more similar to self antigens than they are foreign, and shapes that are more foreign than they are similar to self antigens. The first mode of action of the vaccine is the immune response to parts of the IVIG that are more foreign than similar to self antigens. This immune response is directed against a broad range of foreign shapes in the IVIG that are similar to the broad range of shapes of HIV circulating in the human population. This is the offense arm of the immune response to the vaccine, and is expected to target a multitude of HIV variants. The second mode of action of the vaccine is a response to those components of the vaccine that are more similar to self antigens than they are to foreign antigens. These components stimulate a new set of Th1 cells, that in turn select a new Ts2 population. Co-selection of the Th1 and Ts2 populations results in the establishment of a new Ts2 centre-pole in the system. The repertoire of Th1 cells being stimulated is very broad, and the new complementary Th1 and Ts2 populations are mutually stabilizing. Due to the indirect couplings of Ts2 to the B2 population, the change in the Ts2 centre-pole occurs together with a change in the anti-anti-self IgG population. The new Th1 population is different from the normal Th1 populations of unimmunized persons, that are susceptible to HIV infection. Hence the immune system evades being infected as a result of this mode of action of the vaccine. Furthermore, the anti-anti-self IgG inhibits anti-foreign clones that have anti-self cross-reactivity. This is the defense aspect of the response.

**Predictions of the theory**

Small doses of IVIG given to humans in immunogenic form via an immunogenic route, for example intra-muscularly, are predicted to induce anti-HIV antibodies. Likewise, a mouse version of the vaccine, namely pooled IgG obtained from multiple strains of mice and given to mice in immunogenic form, via an immunogenic route, is predicted to induce anti-HIV antibodies. This is because the total proteome of mice is similar to the total proteome of humans, and the adaptive immune system of mice is similar to that of humans. Consequently the mouse version of the vaccine is expected to perturb the mouse immune system in a way that is similar to the way that IVIG perturbs the human immune system. The same is true for an analogous macaque version of the vaccine, namely macaque IgG obtained from multiple out-bred macaques, given in immunogenic form via an immunogenic route. The macaque vaccine is predicted to result in the production of anti-SIV and anti-HIV antibodies, and to prevent infection of macaques by SIV. Most importantly, small doses of IVIG in immunogenic form given via an immunogenic route are predicted to prevent infection of humans by HIV.

**Conclusion**

At a time when it is widely acknowledged that new approaches are needed for the development of an effective HIV vaccine, recent progress in the development of the symmetrical network theory has led to a remarkably simple theoretical solution to the problem. Predictions for experiments in mice and macaque monkeys using mouse and macaque versions of the vaccine are presented.

Figure 1. An idiotypic network model of immune system regulation that includes MHC class II, Th1 and Th2 helper T cells, Ts1, Ts2 and Ts3 suppressor T cells and serum IgG antibodies. The centre-pole of the system is the Ts2 population. Co-selection is a process in which the members of two populations with complementarity in their specific receptors mutually select each other. There is co-selection of Ts2 and Th1 cells, Ts1 and Ts2 cells, and Ts2 and Ts3 cells. Th2 cells and IgG secreting B cells are both co-selected with both Ts1 and Ts3 cells.

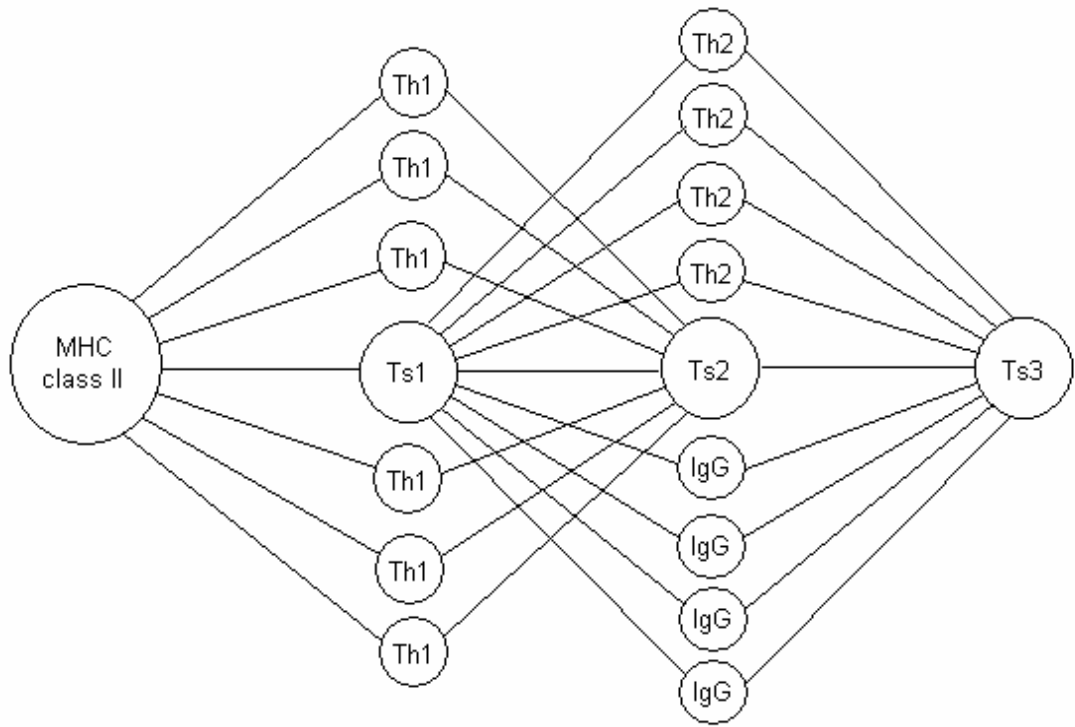

Figure 2. A model showing interactions between all of the self antigens and Th1, Th2, Ts1, Ts2, Ts3, B1 and B2 cells, where B1 cells secrete IgM antibodies and B2 cells secrete IgG antibodies. All the self antigens play a role in the selection of Th1 cells and Ts1 cells, with MHC class II being particularly important. IgM secreting B cells are selected to have complementarity to Ts2. Since in the network model of HIV pathogenesis HIV is selected in the human population to resemble Ts2, normal IgM antibodies have affinity for HIV, and this plausibly accounts for the low level of infectivity of HIV compared with other viruses such as measles.

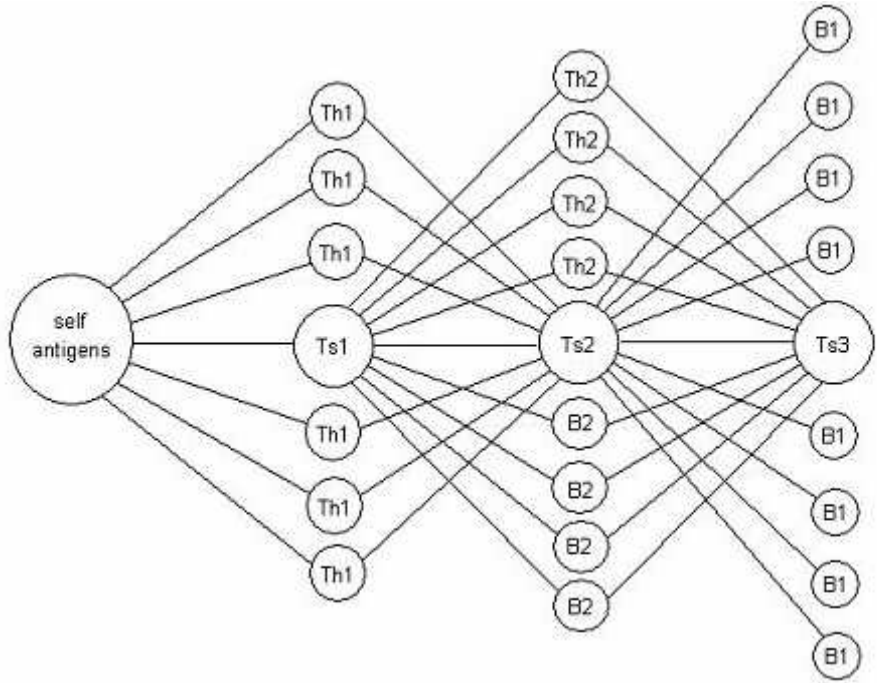